\documentclass{article}

\usepackage{amsmath}
\usepackage{amsthm}
\usepackage{amsfonts}

\begin{document}

\begin{center}
        \Large
        The non-existence of a Lanczos potential for the Weyl curvature
        tensor in dimensions~$n\ge 7$
\end{center}

\begin{center}
        \textbf{
        S. Brian Edgar\footnote{Matematiska institutionen,
                Link\"opings universitet, SE-581 83 Link\"oping, Sweden.
                \label{fn:address}}${}^,$\footnote{E-mail: bredg@mai.liu.se}
        and
        A. H\"oglund${}^{\ref{fn:address},}$\footnote{E-mail:
          anhog@mai.liu.se}
        }
\end{center}

\begin{abstract}
    In this paper it is shown that a Lanczos potential for the
    \emph{Weyl curvature tensor} does not exist
    for all spaces of dimension $n\geq 7$.
\end{abstract}


\

\noindent
Whether there exists a Lanczos potential~\cite{lanczos62} for Weyl
curvature tensors in dimensions $n>4$ has still not been determined.
Although Lanczos's original proof~\cite{lanczos62} for existence was
flawed~\cite{bampi83}, there have subsequently been complete proofs
for existence in \textit{four}
dimensions~\cite{bampi83,illge88,andersson99}.  The latter two
proofs~\cite{illge88,andersson99} are in spinors, although
in~\cite{andersson99} a translation into tensors is given which is
explicitly for \textit{four} dimensions, independent of signature. The
detailed and complicated proof given by Bampi and
Caviglia~\cite{bampi83} is also explicitly for \textit{four}
dimensions, although they also discuss briefly the possibility of
existence in higher dimensions.

An important aspect of all of these proofs is that they are not only
valid for Weyl curvature tensors $C_{abcd}$, but for \textit{Weyl
  candidate tensors} $W_{abcd}$, i.e., any $4$-tensor having the index
symmetries of the Weyl curvature tensor,
\begin{subequations}\label{Wsym}
\begin{equation}\label{Wnocycl}
        W_{abcd} = W_{[ab]cd}  =W_{ab[cd]}=W_{cdab}
        \quad
        W^a{}_{bad}= 0.
\end{equation}
\begin{equation}\label{Wcycl}
        0 = W_{a[bcd]}
\end{equation}
\end{subequations}

In a recent paper~\cite{edgar00} we have shown that a Lanczos
potential for a Weyl candidate tensor does not \textit{generally} exist
for dimensions $n>4$.  In particular, we have shown that in flat and
conformally flat spaces with dimensions $n>4$, the assumed existence
of a Lanczos potential for an arbitrary Weyl candidate imposes
non-trivial conditions on the Weyl candidate.  However, this result
does not say anything about the existence of a Lanczos potential for a
\textit{Weyl curvature tensor} in spaces with non-zero conformal
curvature in dimensions $n>4$.

In this paper we will address the problem of Weyl curvature tensors
\textit{directly}, and show explicitly, that

\

\textit{A Lanczos potential for the Weyl curvature tensor does not
  exist for all spaces of dimension $n\ge7$}.

\

The  $n$-dimensional generalisation for the  Lanczos
potential of a Weyl candidate is given by~\cite{edgar97}
\begin{equation}\label{weyllanczos}
    W^{ab}{}_{cd}
    =
    2  L^{ab}{}_{[c;d]}
    +2L_{cd}{}^{[a;b]}
    -\frac{4}{(n-2)}\delta^{[a}_{[c}\left(
        L^{b]}{}^i{}_{d];i}
        +L_{d]i}{}^{b];i} \right)
\end{equation}
where the Lanczos tensor $L_{abc}$ has the properties that
\begin{subequations}
\begin{equation}
    L_{abc} = L_{bac}
\end{equation}
\begin{equation}\label{eq:gauge}
    L_{ab}{}^b = 0
\end{equation}
\begin{equation}\label{Lcycl}
    L_{[abc]} = 0.
\end{equation}
\end{subequations}
It is easy to check that $W_{abcd}$ satisfies the defining
equations~\eqref{Wsym} for the Weyl candidate tensor.

The condition~\eqref{eq:gauge} is called the Lanczos algebraic gauge
and can always be assumed without loss of generality.  However,
equation~\eqref{weyllanczos} has a more complicated appearance when
the Lanczos algebraic gauge is not imposed.

By checking integrability conditions, it was shown in~\cite{edgar00}
that --- \textit{in dimensions $n\ge 6$} --- \eqref{weyllanczos} leads
to the long complicated condition
\begin{equation}\label{restrictionlong}
\begin{split}
        W{}^{[ab}&{}_{[cd;e]}{}^{f]}
         =
           L_{[cd}{}^{[a;|i|} C^{bf]}{}_{e]i}
        -  L_{[cd}{}^{i;[a} C^{bf]}{}_{e]i}
        -2 L_{[c}{}^{i[a}{}_{;d} C^{bf]}{}_{e]i}
        -2 L_{[c}{}^{i[a;b} C^{f]}{}_{|i|de]}
\\&
        -2 L^{[a|i|}{}_{[c}{}^{;b} C^{f]}{}_{|i|de]}
        +  L_{[c}{}^{i[a} C^{bf]}{}_{de];i}
        +  L^{[a|i|}{}_{[c} C^{bf]}{}_{de];i}
\\&
        -\frac{2}{n-4}\delta^{[a}_{[c}\biggl(
                   L_{de]}{}^{|i;j|} C^{bf]}{}_{ij}
                +2 L_d{}^{|i|b;|j|} C^{f]}{}_{|j|e]i}
                +2 L^{b|i|}{}_d{}^{;|j|} C^{f]}{}_{|i|e]j}
\\&\qquad
                +  L_d{}^{|ij|}{}_{;e]} C^{bf]}{}_{ij}
                +  L^{b|ij|;f]} C_{de]ij}
                +  L_d{}^{|ij|}{}_{;j} C^{bf]}{}_{e]i}
                +  L^{b|ij|}{}_{;|j|} C^{f]}{}_{|i|de]}
\\&\qquad
                +  L_d{}^{|ij|}{}_{;i} C^{bf]}{}_{e]j}
                +  L^{|ij|}{}_d{}^{;b} C^{f]}{}_{e]ij}
                -  L^{|ij|b}{}_{;d} C^{f]}{}_{e]ij}
                +  L^{|ij|b}{}_{;|i|} C^{f]}{}_{|j|de]}
\\&\qquad
                -  L^{|ij|b;f]} C_{de]ij}
                +  L_d{}^{|ij|} C^{bf]}{}_{e]j;i}
                +  L^{b|ij|} C_{de]}{}^{f]}{}_{j;i}
                -\frac{1}{2} L^{|ij|}{}_d C^{bf]}{}_{|ij|;e]}
\\&\qquad
                -\frac{1}{2} L^{|ij|b} C_{de]ij}{}^{;f]}
                +\frac{2}{n-3} L_d{}^{|i|b} C^{f]}{}_{|i|e]j;}{}^{j}
                +\frac{2}{n-3} L^{b|i|}{}_d C_{e]i}{}^{f]}{}_{j;}{}^{j}
        \biggr)
\\&
        -\frac{2}{(n-3)(n-4)} \delta^{[a}_{[c}\delta^b_d \biggl(
                 2 L_{e]}{}^{|ij;k|} C^{f]}{}_{jki}
                +2 L^{f]ij;k} C_{|ijk|e]}
                -  L^{|ij|}{}_{e]}{}^{;|k|} C_{ijk}{}^{f]}
\\&\qquad
                -2 L^{|ij|f];k} C_{|ijk|e]}
                +  L^{|ijk|}{}_{;e]} C_{ijk}{}^{f]}
                +2 L^{|ijk|;f]} C_{|ijk|e]}
                +  L^{|ijk|} C_{|ijk|e]}{}^{;f]}
\\&\qquad
                +  L^{|ijk|} C_{|ijk|}{}^{f]}{}_{;e]}
                +\frac{2}{n-3} L_{e]}{}^{|ij|} C^{f]}{}_{ijk;}{}^{k}
                +\frac{2}{n-3} L^{f]ij} C_{e]ijk;}{}^{k}
\\&\qquad
                +\frac{n-2}{n-3} L^{|ij|}{}_{e]} C_{ij}{}^{f]}{}_{k;}{}^{k}
                +\frac{n-2}{n-3} L^{|ij|f]} C_{|ij|e]k;}{}^{k}
        \biggr)
\\&
        +\frac{4}{(n-3)^2(n-4)} \delta^{[a}_{[c}\delta^b_d\delta^{f]}_{e]}
                L^{ijk} C_{ijkl;}{}^{l}
\\&
        +\frac{1}{(n-2)(n-3)(n-4)} \delta^{[a}_{[c}\delta^b_d\delta^{f]}_{e]}
                C_{ijkl} W^{ijkl}
\\&
        +\frac{1}{n-4} \delta^{[a}_{[c} W^{bf]}{}_{de];i}{}^i
        +\frac{2}{n-4} \delta^{[a}_{[c} W^{|i|b}{}_{de];i}{}^{f]}
        +\frac{2}{n-4} \delta^{[a}_{[c} W^{bf]}{}_{|i|d;e]}{}^i
\\&
        +\frac{4}{(n-3)(n-4)}\delta^{[a}_{[c}\delta^b_d W^{f]i}{}_{e]j;i}{}^j
        -\frac{2}{n-2}W^{[a}{}_{[cd}{}^b \tilde{R}^{f]}{}_{e]}
\\&
        -\frac{4}{(n-2)(n-4)}\delta^{[a}_{[c}W^{b|i|f]}{}_{d}\tilde{R}_{e]i}
        -\frac{4}{(n-2)(n-4)}\delta^{[a}_{[c}W^{b}{}_{de]}{}^{|i|}\tilde{R}^{f]}{}_i
\\&
        -\frac{4}{(n-2)(n-3)(n-4)}\delta^{[a}_{[c}\delta^b_dW^{f]i}{}_{e]}{}^j\tilde{R}_{ij}
\end{split}
\end{equation}
where $\tilde{R}_{ab}$ is the trace free Ricci tensor.  In conformally
flat spaces (i.e., $C_{abcd}=0$) all the terms explicitly containing
$L_{abc}$ disappear, and we obtain, in general, a nontrivial effective
restriction on the Weyl candidate $W_{abcd}$.  However, when we
specialise the Weyl candidate $W_{abcd}$ to the Weyl curvature tensor
$C_{abcd}$ then this restriction collapses in conformally flat spaces.

Let us now consider a particular subclass of $n\ge 7$ dimensional
spaces (where we now use coordinate index notation and let lower case
Latin letters range from 1 to $n$, lower case Greek letters from 1 to
4 and capital Greek letters from 5 to $n$) with the metric
\begin{equation}\label{metric}
    \begin{split}
        ds^2
        &=
        g_{ab}dx^a dx^b
        \\&
        =
        g_{\alpha\beta}dx^\alpha dx^\beta
        + (dx^5)^2+(dx^6)^2+(dx^7)^2+ \ldots +(dx^n)^2
        \\&
        =
        g_{\alpha\beta}dx^\alpha dx^\beta
        + \eta_{\Sigma\Omega} dx^\Sigma dx^\Omega
    \end{split}
\end{equation}
where $g_{\alpha\beta}$ is a \textit{Ricci flat} 4-dimensional metric,
i.e., $g_{\alpha\beta}$ depends only on $x^1$, ..., $x^4$ and its four
dimensional Ricci tensor is zero.

The following properties follow,
\begin{itemize}
\item
    the $n$-dimensional space is Ricci flat, i.e., $R_{ab}=0$, and so
    $R_{abcd} = C_{abcd}$ and from the Bianchi identities
    $C_{abc}{}^d{}_{;d} = 0$.

\item
    all Weyl tensor coordinate components with at least one entry
    of $5,6,7, ..., n$ are zero, i.e.,
    \begin{equation}
        C_{\Sigma bcd}
        = 0 =
        C^{\Sigma}{}_{bcd},
    \end{equation}
    together with all other components from symmetry properties and
    index raising by metric.

\item
    all Christoffel symbols with at least one entry of
    $5,6,7, ... n$ are zero, i.e.,
    \begin{equation}
        \Gamma^{a}_{b\Sigma}
        = \Gamma^{\Sigma}_{ab}
        = 0
    \end{equation}
    together with all other components from symmetry and metric
    properties.
    
\item
    all derivatives of Weyl tensor whose coordinate
    components have at least one entry of $5,6,7, \dots, n$ are zero,
    i.e.,
    \begin{equation}
        C_{\Sigma bcd;e}
        = C_{\Sigma bcd;ef}
        = \dots
        = 0 =
        C^{\Sigma}{}_{bcd;e} \dots
        ,
    \end{equation}
    together with all other components from symmetry properties.
\end{itemize}

When we use this $n$-dimensional metric with the substitutions
$(a,b,f)=(5,6,7)=(c,d,e)$ and $W_{ijkl}=C_{ijkl}$ the
constraint~\eqref{restrictionlong} simplifies a lot because almost all
of the products contain a term with $C_{\Sigma bcd}$.
So~\eqref{restrictionlong} becomes
\begin{equation}
    C_{ijkl} C^{ijkl}=0.
\end{equation}

This is a restriction on the metric~\eqref{metric}, which translates
directly to the 4-dimensional Ricci-flat metric $g_{\alpha \beta}$;
since it requires a zero value for a normally non-zero Riemann scalar
invariant of this metric, (even in Ricci flat case, e.g.,
Schwarzschild) it is therefore a non-trivial effective restriction.

If we do the same, but without setting $W_{abcd} = C_{abcd}$, we get a
restriction on the \textit{Weyl candidate} tensor,
\begin{equation}\label{restrictgeneral}
\begin{split}
        W{}^{[ab}{}_{[cd;e]}{}^{f]}
        & =
        \frac{1}{(n-2)(n-3)(n-4)} \delta^{[a}_{[c}\delta^b_d\delta^{f]}_{e]}
                C_{ijkl} W^{ijkl}
\\&
        +\frac{1}{n-4} \delta^{[a}_{[c} W^{bf]}{}_{de];i}{}^i
        +\frac{2}{n-4} \delta^{[a}_{[c} W^{|i|b}{}_{de];i}{}^{f]}
\\&
        +\frac{2}{n-4} \delta^{[a}_{[c} W^{bf]}{}_{|i|d;e]}{}^i
        +\frac{4}{(n-3)(n-4)}\delta^{[a}_{[c}\delta^b_d W^{f]i}{}_{e]j;i}{}^j
\end{split}
\end{equation}
This is an effective restriction since $W_{abcd} = C_{abcd}$ does not
satisfy it.  It is more general than it first appears to be; in the
derivation the conditions~\eqref{Wcycl} and~\eqref{Lcycl} have not
been used.  Condition~\eqref{restrictgeneral} is therefore also a
restriction on the existence of a Lanczos potential for the larger
class of Weyl candidate tensors which lack the symmetry~\eqref{Wcycl}.
The existence problem without the conditions~\eqref{Wcycl}
and~\eqref{Lcycl} is referred to as the \textit{parallel problem}
in~\cite{edgar00} and~\cite{bampi83}.

\

Finally we emphasise that although condition~\eqref{restrictionlong}
is also applicable to spaces of dimension six, when we construct an
analogous metric to~\eqref{metric} it does not have the same crucial
properties in $6$-dimensional spaces, and so we cannot draw the same
conclusions as above.  As regards $5$-dimensional spaces, we showed
in~\cite{edgar00} that condition~\eqref{restrictionlong} is trivially
satisfied, but that another --- even more complicated --- condition
applies (involving third derivatives of the Lanczos potential, and too
complicated to write out explicitly).  So the question of existence of
a Lanczos potential specifically for the Weyl curvature tensor in all
spaces of dimensions five and six has still not been formally ruled
out.

\section*{Acknowledgements.}

SBE wishes to acknowledge the ongoing financial support of
V.R.~(Swedish Natural Sciences Research Council).


\begin{thebibliography}{9}

\bibitem{lanczos62}
    Lanczos, C. (1962).
    ``The splitting of the Riemann tensor''
    \textit{Rev. Mod. Phys.}, \textbf{34}, 379--389.

\bibitem{bampi83}
    Bampi, F., and Caviglia, G. (1983).
    ``Third-order tensor potentials for the Riemann and Weyl tensors'',
    \textit{Gen. Rel. Grav.}, \textbf{15}, 375--386.

\bibitem{illge88}
    Illge, R. (1988).
    ``On potentials for several classes of spinor and tensor fields in
    curved spacetimes'',
    \textit{Gen. Rel. Grav.}, \textbf{20}, 551--564.

\bibitem{andersson99}
    Andersson, F. and Edgar S. B. (2001).
    ``Existence of Lanczos potentials and superpotentials for the Weyl
    spinor/tensor'',
    \textit{Class. Quantum. Grav}, \textbf{18}, 2297--2304.

\bibitem{edgar00}
    Edgar, S. B. and  H\"oglund, A. (2000).
    ``The Lanczos potential for Weyl-candidate tensors exists only in
    four dimensions'',
    \textit{Gen. Rel. Grav.}, \textbf{32},  2307--2318,
    gr-qc/9711022.

\bibitem{edgar97}
    Edgar, S. B. and  H\"oglund, A. (1997).
    ``The Lanczos potential for the Weyl curvature tensor; existence,
    wave equation and algorithms'',
    \textit{Proc. Roy. Soc. London Ser.}, A \textbf{453}, 835--851,
    gr-qc/9601029.

\end{thebibliography}
\end{document}